\documentclass[preprint,
superscriptaddress,
 amsmath,amssymb,
 aps,
 prb,
 floatfix,
 10pt,
 twocolumn]
{revtex4-2}

\usepackage{graphicx} 
\usepackage{dcolumn}
\usepackage{color}
\usepackage{xcolor,colortbl}
\usepackage{bm}

\usepackage{subcaption}
\usepackage{multirow}

\usepackage{xcolor}

\begin{document}

\title{Long-range machine-learning potentials with environment-dependent charges enable predicting LO-TO splitting and dielectric constants}

\author{Dmitry Korogod}
\email{korogod.dv@phystech.su}
\affiliation{Skolkovo Institute of Science and Technology, Skolkovo Innovation Center, Bolshoy boulevard 30, Moscow, 121205, Russian Federation}
\affiliation{Moscow Institute of Physics and Technology, 9 Institutskiy per., Dolgoprudny, Moscow Region, 141701, Russian Federation}
\affiliation{Digital Materials LLC, Odintsovo, Kutuzovskaya str. 4A Moscow region, 143001, Russian Federation}

\author{Alexander V. Shapeev}
\affiliation{Skolkovo Institute of Science and Technology, Skolkovo Innovation Center, Bolshoy boulevard 30, Moscow, 121205, Russian Federation}
\affiliation{Digital Materials LLC, Odintsovo, Kutuzovskaya str. 4A Moscow region, 143001, Russian Federation}

\author{Ivan S. Novikov}
\affiliation{HSE University, Faculty of Computer Science, Pokrovsky boulevard 11, Moscow, 109028, Russian Federation}

\date{\today}

\begin{abstract}

We present two models with explicit long-range electrostatics in the form of Coulomb interactions. Both models include point charges depending on their local atomic environments, and the second model also conserves a total charge of an atomic system. We combine the proposed long-range models with local Moment Tensor Potential and demonstrate that they reduce the training errors of the MTP models fitted on the same training sets including the CH$_3$COO$^-$+4-methylphenol and CH$_3$COO$^-$+4-methylimidazole organic dimers (non-periodic systems) and the NaCl crystal (periodic system). For the organic dimers, the proposed models also give qualitatively correct predictions of the binding curves. Furthermore, in this study we introduce a method for calculating phonon spectra of isotropic materials only via these long-range models fitted to energies, forces, and stresses. The developed long-range model with point charges dependent on atomic environments and conserving total charge is capable of predicting the correct value of the LO-TO splitting in the $\Gamma$-point in the isotropic NaCl. For this system, we also predict dielectric constant from dipole moment fluctuations calculated with molecular dynamics simulations conducted with the developed long-range model. The calculated dielectric constant is in good agreement with experiment. Finally, we demonstrate the broader applicability of the introduced approach by computing the phonon spectrum of uniaxial tetragonal PbTiO$_3$. Although the method is formally derived for isotropic materials, we show that it is also perspective for uniaxial materials (e.g., PbTiO$_3$) as the spectrum obtained with our long-range interatomic potential corresponds to the one calculated with density functional theory.
    
\end{abstract}

\maketitle

\section{Introduction}

Machine-learning interatomic potentials (MLIPs) explicitly incorporating long-range electrostatic interactions in their functional form have proven to be a reliable tool in computational materials science. These potentials enable improving accuracy of short-range (local) interatomic potentials. This fact was demonstrated, e.g., on the example of water: in~\cite{DPLR}, the DPLR model, which is the combination of the Deep Potential Molecular Dynamics (DeePMD) model and the long-range model incorporating two different types of charged sites, improved the prediction of the interaction energy of water dimer and of the free energy profile of the water molecule absorption to a water slab, and in~\cite{LatentEwald}, Latent Ewald Summation (LES) was combined with Cartesian Atomic Cluster Expansion (CACE) and the accuracy of the short-range CACE model was improved with the CACE+LES model for predicting the distribution of water molecule orientations at the liquid-gas interface. More details on the above and other models that explicitly incorporate long-range interactions are given in~\cite{korogod2025_mtp_coulomb_fixed}.

In this study, we propose two simple models with explicit long-range electrostatic interactions in the form of the Coulomb interaction of point charges. The first model is called Environment-Dependent Charges (EDQ) and includes atomic charges dependent only on their local atomic environments. The second model is Environment-Dependent Charge Redistribution (EDQRd), which inherits features of the EDQ model, namely, charges dependent on their environments, and features of the Charge Redistribution (QRd) model proposed in ~\cite{korogod2025_mtp_coulomb_fixed}, which provides computationally cheap total charge conservation. We combine each of these long-range models with the short-range Moment Tensor Potential~\cite{MTPsingle,MTPmulti} and compare these combined models. We note that the proposed models with atomic charges depending on local atomic environments can be directly combined with any other short-range MLIP.

We show the importance of taking the dependence of charges on their local environments into account on examples of molecular (non-periodic) systems: CH$_3$COO$^-$+4-methylphenol and $\rm CH_3COO^-$+4-methylimidazole. In particular, the MTP+EDQ model gives three times smaller energy errors than MTP+QRd and the MTP+EDQ model is capable of correctly reproducing the binding curve of $\rm CH_3COO^-$+4-methylimidazole predicted with density functional theory (DFT) as opposed to MTP+QRd.

The main results and tests of this work are related to the applicability of the MTP+EDQRd model to periodic systems. We first investigate the NaCl crystal and demonstrate that MTP+EDQRd gives energy, force, and stress fitting errors several times smaller than the MTP model fitted on the same training set created with active learning of MTP~\cite{podryabinkin2017_AL} during molecular dynamics (MD) simulations. The density and lattice constant predicted with these models are in good agreement with DFT and experiment. An important result of this study is a proposed method for calculating phonon spectra of isotropic materials. Our findings show that it is not necessary to know neither the high-frequency dielectric constant nor Born Effective Charges (BECs) typically obtained with DFT for calculating a non-analytical correction (NAC) to the dynamical matrix used to capture the LO-TO splitting in the $\Gamma$-point. To illustrate this fact, we calculated this NAC solely via point charges and their derivatives predicted with MTP+EDQRd. The phonon spectrum obtained with our model is in good correspondence with the DFT one. For the isotropic NaCl, we also predict the ratio of the static and high-frequency dielectric constants from fluctuations of the dipole moment using MTP+EDQRd and demonstrate its correspondence to the experiment. 

Finally, we explore the applicability of the MTP+EDQRd model to the more challenging PbTiO$_3$ crystal. We use training and validation sets from~\cite{PbTiO3_training_set}, fit MTP+EDQRd, and obtain close fitting and validation errors to the recently developed CACE+LES model. The broader applicability of the introduced method for phonon spectra calculations is also demonstrated. Although the proposed approach is strictly valid only for isotropic crystals, the tetragonal PbTiO$_3$ phonon spectrum predicted only via the MTP+EDQRd model corresponds to the DFT one.

\section{Methodology}

\subsection{Moment Tensor Potential}

As a short-range (local) model we use MTP~\cite{MTPsingle,MTPmulti}, which has already been applied to the investigation of binding curves in combination with the Coulomb model including fixed charges in~\cite{korogod2025_mtp_coulomb_fixed}. The energy of MTP is a sum of contributions $V_i$ of local environments:
\begin{equation} \label{MTPenergy}
   E^{\rm short}\left(\bm x, \bm \theta \right) \equiv E^{\rm MTP}\left(\bm x, \bm \theta \right) = \sum_{i=1}^N V_i\left(\bm \theta \right) = \sum_{i=1}^N V\left(\mathfrak{n}_i, \bm \theta \right),
\end{equation}
where $\bm x$ is a configuration, $N$ is a number of atoms in it, $\bm \theta$ is a vector of potential parameters to be fitted, and $\mathfrak{n}_i=\{\bm r_{ij}, z_i, z_j\}$ is a local environment of the $i$-th atom including the relative positions $\bm r_{ij} = \bm r_i - \bm r_j$ between the central atom $i$ and its neighboring atoms (neighbors) $j$ and their atomic types $z_i$ and $z_j$, respectively. The number of neighbors in the atomic environment is determined by the cutoff radius $R_{\rm cut}$: only atoms with distance from the central $i$-th atom $r_{ij} = \left\vert \bm r_{ij} \right\vert < R_{\rm cut}$ are considered as its neighbors.

In MTP, we determine $V_i$ as a linear combination of basis functions $B_{\alpha}$:
\begin{equation} \label{MTPsite_en}
    V \left(\mathfrak{n}_i, \bm \theta \right) = \zeta_{z_i} + \sum_{\alpha} \xi_{\alpha} B_{\alpha}
    \left( \mathfrak{n}_i, \hat{C} \right),
\end{equation}
where $\bm \xi$ are the linear parameters of the model, $\bm \zeta$ are the free parameters of the model, and $\hat{C}$ are the so-called radial parameters of the model. These three parameter sets constitute all the learnable parameters of MTP: $\bm \theta = \left(\bm \zeta, \bm \xi, \hat{C} \right)$.

Basis functions $B_{\alpha}(\mathfrak{n}_i, \hat{C})$ are determined as all possible contractions (or, products) of the so-called moment tensor descriptors
\begin{equation} \label{MTD}
M_{\mu,\nu}(\mathfrak{n}_i, \hat{C}) = \sum_j \sum_{\beta=1}^{\beta_{\rm max}} c_{\mu,z_i,z_j}^{(\beta)} T^{(\beta)}(r_{ij})(R_{\rm cut} - r_{ij})^2 \bm r_{ij}^{\otimes \nu} 
\end{equation}
yielding a scalar. In \eqref{MTD}, $\mu$ is a radial number of the moment tensor descriptor, $\beta$ is an order of Chebyshev polynomial $T^{(\beta)}(r_{ij})$, $\beta_{\rm max}$ is the radial basis size, $c_{\mu,z_i,z_j}^{(\beta)}$ are the elements of $\hat{C}$, and a symbol ``$\otimes$'' is an outer product of vectors, and, thus, $M_{\mu,\nu}$ is the tensor of the $\nu$-th order. 

A number of contractions of moment tensor descriptors yielding a scalar is infinite. To restrict this number for the construction of a functional form of MTP that can be used during calculations, we introduce the so-called level of moment tensor descriptors ${\rm lev} M_{\mu, \nu} = 2 + 4 \mu + \nu$. Next, if the product $\prod_{s=1}^S M_{\mu_s, \nu_s}$ forms a basis function $B_{\alpha}$ then ${\rm lev} B_{\alpha} = 2S + \sum_{s=1}^S (4 \mu_s + \nu_s)$. Finally, we introduce the so-called level of MTP ${\rm lev}_{\rm MTP}$ and include in its functional form only the basis functions with ${\rm lev} B_{\alpha} \leq {\rm lev}_{\rm MTP}$. 

\subsection{Long-range models}

To obtain a long-range model, we combine two independent contributions: the short-range energy $E^{\rm short}$ predicted by the local potential and the electrostatic energy $E^{\rm elec}$ predicted by an electrostatic model, giving the total energy
\begin{equation}
    \label{eq:tot_en}
    E^{\rm long}\left(\bm x, \bm \theta, \bm a \right) = E^{\rm short}\left(\bm x, \bm \theta \right) + E^{\rm elec}\left(\bm x, \bm a \right),
\end{equation}
where $\bm \theta$ and $\bm a$ are the independent parameters of the short-range and electrostatic parts, respectively. We note that hyperparameters of the short-range and long-range parts, e.g. cutoff radius $R_{\rm cut}$, the radial basis size $\beta_{\rm max}$, or the level of MTPs from which these parts are composed, are also independent and can be set up different for calculations.

We use three electrostatic models, which have one thing in common: the electrostatic energy in them is given by the Coulomb interaction of point charges centered on atoms:
\begin{equation}
    \label{eq:LREnergy}
    E^{\rm elec} = \sum_{j<i}\frac{q_iq_j}{r_{ij}},
\end{equation}
$q_i$ and $q_j$ are the charges of the $i$-th and $j$-th atoms, respectively, and the Ewald summation~\cite{ClassicalEwaldSummation} is used to calculate this sum for the systems in periodic boundary conditions. The ways in which these charges are predicted are different for each model and are described in the next subsections.

\subsubsection{Charge Redistribution}

In the first model called Charge Redistribution~(QRd) and introduced in~\cite{korogod2025_mtp_coulomb_fixed}, atomic charges depend only on the chemical composition of a system (i.e., on the vector of atomic types $\bm z$ in a system) and sum up to the total charge of a system, which ensures conservation of the total charge, including non-zero:
\begin{equation}
    \label{eq:charges_qrd}
    q_i({\bm z}, {\bm a}) = q_i({\bm z}, ({\bm b}, {\bm s})) = b_{z_i} + s_{z_i} \frac{Q_{\rm total} - \sum_{j=1}^N b_{z_j}}{\sum_{j=1}^N s_{z_j}}.
\end{equation}
In this equation, $\bm b$ and $\bm s$ are the vectors of model parameters, $Q_{\rm total}$ is a total charge of a system defined in the beginning of the simulation.

\subsubsection{Environment-Dependent Charges}

In the second model called Environment-Dependent Charges~(EDQ), atomic charges depend only on local environments $\mathfrak{n}_i$:
\begin{equation}
    \label{eq:charges_edq}
    q_i=q\left(\mathfrak{n}_i, \bm a \right)=V\left(\mathfrak{n}_i, \bm a \right).
\end{equation}
While any local model can be utilized to predict charges in EDQ, in this work we use MTP. We note that despite the fact that this model is naturally much more accurate than the QRd model, it lacks total charge conservation. Therefore, EDQ should be used only when it is not crucial: for example, for simulations in vacuum.

\subsubsection{Environment-Dependent Charge Redistribution}

The third model is called Environment-Dependent Charge Redistribution~(EDQRd) and combines the main ideas of QRd and EDQ. In this model, atomic charges are predicted on the basis of the local environments of all atoms and on the chemical composition of the system:
\begin{equation}
	\label{eq:charges_edqrd}
    \begin{split}
    	q_i\left(\bm x, \bm a\right) &= q_i\left(\bm x, (\bm p, \bm s)\right) \\= V\left(\mathfrak{n}_i, \bm p\right)
        &+s_{z_i}\frac{Q_{\rm total} - \sum_{j=1}^N V\left(\mathfrak{n}_j, \bm p\right)}{\sum_{j=1}^N s_{z_j}},
    \end{split}
\end{equation}
where $\bm p$ is a vector of parameters of a short-range model used for charge prediction.

EDQRd can be seen as QRd in which the type-dependent parameter $\bm b$ was substituted by some environment-dependent term. This model combines the accuracy of EDQ with computationally cheap charge conservation of QRd. Note that the computational cost of calculating charges of all atoms in the system is the same as the cost of computing energy of a system using underlying short-range model since the redistribution part is computationally cheap.

\subsection{Fitting}

Assume that we have a training set with $K$ configurations and DFT energies $E^{\rm DFT}\left(\bm x_{(k)}\right)$, forces $F^{\rm DFT}_{i, l}\left(\bm x_{(k)}\right)$, and stresses $\sigma^{\rm DFT}_{a,b}\left(\bm x_{(k)}\right)$ for each configuration $\bm x_{(k)}$ $\big(k=1,\ldots,K$; $l,a,b=1,2,3\big)$. To train a MLIP predicting energy $E\left(\bm x, \bm \Omega\right)$, forces $F_{i, l}\left(\bm x, \bm \Omega \right)$, and stresses $\sigma_{a,b}\left(\bm x, \bm \Omega\right)$ for a configuration $\bm x$, we minimize the loss function
\begin{equation}
		\label{eq:lossfunction}
		\begin{split}
            &\mathcal{L}(\bm \Omega) = \sum_{k=1}^K {\Biggl\lbrack}
            w_e \left(E\left(\bm x_{(k)}, \bm \Omega\right)-E^{\rm DFT}\left(\bm x_{(k)}\right)\right)^2\\
            &+
            w_f \sum_{i=1}^{N_k}  \sum_{l=1}^3 \left(F_{i, l}\left(\bm x_{(k)}, \bm \Omega\right)-F^{\rm DFT}_{i, l}\left(\bm x_{(k)}\right)\right)^2\\
            &+w_s \sum_{a=1}^3 \sum_{b=1}^3 \left(\sigma_{a, b}\left(\bm x_{(k)}, \bm \Omega\right)-\sigma^{\rm DFT}_{a, b}\left(\bm x_{(k)}\right)\right)^2
    {\Biggr\rbrack}
			\end{split}
\end{equation}
with respect to the parameters $\bm \Omega$ of the model. In~\eqref{eq:lossfunction}, $w_e$, $w_f$, and $w_s$ are non-negative weights, determining the importance of energies, forces, and stresses with respect to each other. In this work, we performed this minimization using the Broyden--Fletcher--Goldfarb--Shanno (BFGS) algorithm. We note that here we fit each type of model by minimizing \eqref{eq:lossfunction}.

\subsection{Active Learning}

To automatically construct a training set for fitting the long-range models for NaCl, we used the active learning algorithm proposed for the short-range MTP model in~\cite{podryabinkin2017_AL}. As it is demonstrated in Section~\ref{sec:results}, a training set created in this way enables fitting a reliable and accurate long-range model.

Denote a number of configurations in an initial training set by $K$. For each configuration $\bm x_{(k)}, k=1,\ldots,K$ we have the reference energy $E_k^{\rm DFT}$ obtained using DFT, and we found the vector of MTP optimal parameters $\bm \theta^*$ of length $m$ by minimizing 
$$\sum_{k=1}^K\left( E_k^{\rm MTP}\left(\bm \theta\right) - E_k^{\rm DFT}\right)^2$$
with respect to $\bm \theta$. Here $E_k^{\rm MTP}(\bm \theta) \equiv E^{\rm MTP}\left(\bm x_{(k)}, \bm \theta\right)$. Then, we can reformulate the problem of fitting MTP as the problem of solving an overdetermined system of equations with respect to $\theta_p$
\begin{align} \label{eq:OverdeterminedSystem}
\sum \limits_{p=1}^{m} \theta_p \dfrac{\partial E^{\rm MTP}_k(\bm \theta^*)}{\partial \theta_p} = E^{\rm DFT}_k + \sum \limits_{p=1}^{m} \theta^*_p \dfrac{\partial E^{\rm MTP}_k(\bm \theta^*)}{\partial \theta_p}
\end{align}
with the matrix
\begin{equation}
\label{eq:MatrixB}
    B = \begin{pmatrix}
\frac{\partial E_1^{\rm MTP}}{\partial \theta_1} (\bm \theta^*) & \dots & \frac{\partial E_1^{\rm MTP}}{\partial \theta_m} (\bm \theta^*) \\
\vdots & \ddots & \vdots \\
\frac{\partial E_K^{\rm MTP}}{\partial \theta_1} (\bm \theta^*) & \dots & \frac{\partial E_K^{\rm MTP}}{\partial \theta_m} (\bm \theta^*)
\end{pmatrix},
\end{equation} 
in which each row corresponds to the atomic configuration in the training set. The maxvol algorithm \cite{goreinov2010_maxvol} enables finding the $m \times m$ submatrix $A$ that maximizes the absolute value of the determinant (volume) from the matrix $B$ of size $K \times m$. In other words, we find $m$ most linearly independent rows in the matrix $B$ which correspond to $m$ most geometrically different configurations. 

Once the matrix $A$ is constructed, we start any atomistic simulation, e.g., molecular dynamics as for NaCl in this study, with the so-called extrapolation control. To that end, for each configuration encountered during a simulation, we calculate the extrapolation grade $\gamma$:
\begin{equation}
\begin{array}{c}
	\label{eq:grade}
	\gamma=\max_{1 \leq j \leq m} |c_j|, \\ 
	~\bm c = \left(
	\frac{\partial E^{\rm MTP}}{\partial\theta_1}\left(\bm \theta^* \right), \dots, \frac{\partial E^{\rm MTP}}{\partial\theta_m}\left(\bm \theta^* \right)
	\right)A^{-1}.
\end{array}
\end{equation}
This value determines a factor by which $|{\rm det}(A)|$ changes if we substitute the $k$-th row in $A$ with the row $\left(
\frac{\partial E^{\rm MTP}}{\partial\theta_1}\left(\bm \theta^* \right), \dots, \frac{\partial E^{\rm MTP}}{\partial\theta_m}\left(\bm \theta^*\right)
\right)$ where $k = \arg \max_{j} |c_j|$. If $\gamma < 1$ then the volume of the matrix $A$ decreases in the case of adding this configuration to this matrix, and the current potential interpolates on this configuration. Otherwise, the volume of this matrix increases and we call that the potential extrapolates on this configuration. To formulate the active learning algorithm, we introduce two thresholds: $\gamma_{\rm th}$ and $\Gamma_{\rm th}$, $1<\gamma_{\rm th} \leq \Gamma_{\rm th}$. The thresholds $\gamma_{\rm th}$ and $\Gamma_{\rm th}$ are the lower and upper bounds of the permissible (reliable) extrapolation for which the current potential is slightly off in predicting energy, forces, and stresses of the current configuration and we can continue atomic simulations with the current potential. However, if $\gamma_{\rm th} \leq \gamma \leq \Gamma_{\rm th}$ then the extrapolation is large enough, the current configuration is a candidate to be added to the training set, and we add it to the preselected set. If $\gamma > \Gamma_{\rm th}$ then the extrapolation is risky and further atomistic simulation with the current potential may lead to non-physical configurations and, therefore, we break an atomistic simulation and add the current configuration to the preselected set. 

Assume we have $P$ preselected configurations. In the next step of the active learning algorithm, we add the rows $\left(
\frac{\partial E_p^{\rm MTP}}{\partial\theta_1}\left(\bm \theta^* \right), \dots, \frac{\partial E_p^{\rm MTP}}{\partial\theta_m}\left(\bm \theta^*\right)
\right)$, $p=1,\ldots,P$ to the matrix $A$ of size $m \times m$ and obtain the matrix $\tilde{B}$ of size $(P+m) \times m$. We then use the maxvol algorithm and update the matrix $A$. After that, we conduct DFT calculations for all the configurations corresponding to the rows that substituted the rows in the previous matrix $A$ and refer to these configurations as selected. We add the selected configurations to the training set, re-train the current MTP, and update the matrix $A$ again using the re-trained MTP. We then re-start an atomistic simulation, select new configurations to be added to the training set, and re-fit the potential on the updated training set. The active learning algorithm described above is terminated when the number of preselected configurations is equal to zero.

\subsection{Calculating phonon spectra of isotropic materials}

In this subsection, we describe a method for calculating phonon spectra of isotropic materials using long-range interatomic potentials. To account for the contributions of the dipole-dipole interactions in the vicinity of the $\Gamma$-point (which is necessary to capture the LO-TO splitting), a non-analytical correction to the dynamical matrix is added, which is represented by the following additional term in the force constants~$\Phi_{\alpha\beta}^{\rm dd}\left(
0\kappa;j\kappa'\right)$~\cite{PhononsNACDipoleDipole}:
\begin{equation}
    \label{eq:nac}
    \begin{split}
\Phi_{\alpha\beta}^{\rm dd}&\left(
0\kappa;j\kappa'\right) =\sum_{\alpha'\beta'} Z^*_{\kappa, \alpha\alpha'}Z^*_{\kappa', \beta\beta'} \\
&\times\left(\frac{\left(\hat{\varepsilon}_\infty^{-1}\right)_{\alpha'\beta'}}{D^3}-3\frac{\Delta_{\alpha'}\Delta_{\beta'}}{D^5}\right)\left(\operatorname{{det}{\hat{\varepsilon}_\infty}}\right)^{-0.5},
\end{split}
\end{equation}
where $\hat{\varepsilon}_\infty$ is the high-frequency dielectric tensor, and $Z^*_{\kappa, \alpha \alpha'}$ is the $\alpha \alpha'$-component of a Born Effective Charge~(BEC) of the $\kappa$-th atom, where $\Delta_\alpha=\sum_\beta\left(\hat{\varepsilon}_\infty^{-1}\right)_{\alpha\beta}d_\beta$, $\bm d$ is the position of the $\kappa'$-th atom in the $j$-th unit cell relative to the position of the $\kappa$-th atom in the 0-th unit cell, and $D=\sqrt{\sum_\alpha \Delta_\alpha d_\alpha}$.

While usually either DFT calculations or experiments are required to obtain BECs and high-frequency dielectric tensor, in this work we propose a method to calculate a non-analytical correction term for isotropic materials using only atomic charges, predicted by our models.

Suppose that $\left(\hat{\varepsilon}_{\infty}\right)_{\mu \lambda} = \varepsilon_\infty \delta_{\mu \lambda}$ (which is true, for instance, for isotropic materials), where $\delta_{\mu \lambda}$ is the Kronecker delta. Then, substituting this into~\eqref{eq:nac}, we get
\begin{equation}
    \label{eq:nac2}
\Phi_{\alpha\beta}^{\rm dd}\left(
0\kappa;j\kappa'\right) =\sum_{\alpha'\beta'} \frac{Z^*_{\kappa, \alpha\alpha'}Z^*_{\kappa', \beta\beta'}}{\varepsilon_\infty}\left(\frac{\delta_{\alpha'\beta'}}{\vert\rm d\vert^3}-3\frac{d_{\alpha'}d_{\beta'}}{\vert\rm d\vert^5}\right).
\end{equation}
Next, we utilize the following definition of polarization $\bm P$, proposed in~\onlinecite{LES_BEC}:
\begin{equation}
    \label{eq:polarization_from_pac}
        P_\alpha = \sqrt{\varepsilon_\infty}\underbrace{\frac{1}{2\pi i}\sum_\beta R_{\alpha\beta}\sum_{j=1}^Nq_j 
        \operatorname{exp}\left(
        2\pi i \rho_{j, \beta}\right)}_{P_\alpha^0}=\sqrt{\varepsilon_\infty} \, P_\alpha^0,
\end{equation}
where $i$ is an imaginary unit, $\hat{R}$ is a cell matrix, $R_{\alpha\beta}$ is its $\alpha \beta$-component, and $\rho_{j, \beta}$ is $\beta$-component of the reduced coordinates of $j$-th atom~$\left(\rho_{j, \beta}=\sum_\gamma \left(\hat{R}^{-1}\right)_{\beta\gamma}r_{j,\gamma}\right)$. Note that $P_\alpha^0$ is independent of the high-frequency dielectric constant and can be computed solely via our machine-learning potential. We also want to emphasize that all the reasoning behind square root of the high-frequency dielectric constant being in this definition from the original article holds for our models.

Then, the following equation for the BECs is given in~\cite{LES_BEC}:
\begin{equation}
    \label{eq:bec_from_pac}
    \begin{split}
    Z^*_{l, \alpha \beta} = \operatorname{Re}{\left( \exp{\left(
        -2\pi i \rho_{l, \alpha}\right)}\frac{\partial 
    P_\alpha}{\partial \, r_{l, \beta}}\right)}
    \\= \sqrt{\varepsilon_\infty} \underbrace{\operatorname{Re}{\left( \exp{\left(
        -2\pi i \rho_{l, \alpha}\right)}\frac{\partial 
    P_\alpha^0}{\partial \, r_{l, \beta}}\right)}}_{Z^0_{l, \alpha \beta}}
    = \sqrt{\varepsilon_\infty} \, Z^0_{l, \alpha \beta},
    \end{split}
\end{equation}
where $\operatorname{Re}$ is a real part of a complex number, and $Z^0_{l, \alpha \beta}$ is the $\alpha \beta$-component of a scaled BEC of $l$-th atom that is independent of high-frequency dielectric constant and can be calculated using our machine learning model.

Substituting \eqref{eq:bec_from_pac} into \eqref{eq:nac2} gives
\begin{equation}
    \label{eq:nacfinal}
\Phi_{\alpha\beta}^{\rm dd}\left(
0\kappa;j\kappa'\right) =\sum_{\alpha'\beta'} Z^0_{\kappa, \alpha\alpha'}Z^0_{\kappa', \beta\beta'}\left(\frac{\delta_{\alpha'\beta'}}{\vert\rm d\vert^3}-3\frac{d_{\alpha'}d_{\beta'}}{\vert\rm d\vert^5}\right).
\end{equation}
Note that the right-hand side of~\eqref{eq:nacfinal}
is independent of the high-frequency dielectric constant and can be calculated for any given system using our machine-learning potential. We used~\eqref{eq:nacfinal} to calculate the non-analytical correction and compared the obtained phonon spectrum with the DFT data.

\subsection{Calculating the ratio of static and high-frequency dielectric constants of isotropic materials}

Based on the molecular dynamics (MD) in electronic continuum theory~\cite{MDEC}, which inspired the equations for polarization and BECs given in~\cite{LES_BEC} and utilized in the present work, the dielectric constant predicted from dipole moment fluctuations in MD simulations
\begin{equation}
    \label{eq:diel_const}
    \varepsilon_{\rm MD} = 1 + \frac{4\pi}{3V\varkappa T}\left(\left\langle\bm M^2\right\rangle-\left\langle\bm M\right\rangle^2\right)
\end{equation}
is the ratio of static ($\varepsilon_0$) and high-frequency ($\varepsilon_{\infty}$) dielectric constants. In~\eqref{eq:diel_const}, $V$ is a volume of a system, $\varkappa$ is the Boltzmann constant, $T$ is a temperature of a system, and the total dipole moment $\bm M$ of a system is
\begin{equation}
    \label{eq:dip_mom}
    \bm M = \sum_{i=1}^N q_i\bm r_i^u,
\end{equation}
where $\bm r_i^u$ are the so-called unwrapped coordinates of the $i$-th atom, i.e., the coordinates the atom would have had if it was not wrapped back into the simulation cell according to periodic boundary conditions every time it exited the cell. We used~\eqref{eq:diel_const} to calculate the dielectric constant predicted from dipole moment fluctuations and compared the obtained value with the ratio of experimental static and high-frequency constants.

\section{Results and discussion}
\label{sec:results}

\subsection{Improvement of binding curves with MTP+EDQ}

Firstly, we tested MTP+EDQ on two molecular dimers in vacuum (non-periodic systems), namely, $\rm CH_3COO^-$+4-methylphenol and $\rm CH_3COO^-$+4-methylimidazole, for which MTP+QRd was unable to significantly improve accuracy of MTP~\cite{korogod2025_mtp_coulomb_fixed}. Training and test sets were taken from~\cite{korogod2025_mtp_coulomb_fixed}. For $\rm CH_3COO^-$+4-methylphenol, these sets contained 312 and 36 dimer configurations, respectively, and for $\rm CH_3COO^-$+4-methylimidazole they contained 313 and 32 dimer configurations, respectively.

Similarly to what was done in~\cite{korogod2025_mtp_coulomb_fixed}, we used MTP of the 8-th level with eight radial basis functions and 9~\AA \;cutoff as a short-range part. In EDQ, an MTP of the 2-nd level with the same radial basis settings (including the cutoff radius) was utilized as the model predicting charges.

For both systems, ensembles of five MTPs, five MTP+QRd models, and five MTP+EDQ models were trained. We used $w_e = 1.0 \; \left(\text{eV}\right)^{-2}, \;w_f = 0.01 \; \left(\rm eV/\rm \AA\right)^{-2}, \;w_s = 0.0 \; \left(\text{eV}\right)^{-2}$ during the training of both short-range and long-range models. Training errors obtained for these systems are shown in Table~\ref{tab:dimers_errors}.

\begin{table*}[!ht]
\caption{Fitting errors for the $\rm CH_3COO^-$+4-methylphenol and $\rm CH_3COO^-$+4-methylimidazole systems. The results are given for the ensemble of five models with 1-$\sigma$ confidence interval.}
\label{tab:dimers_errors}
\begin{center}
\begin{tabular}{c|c|c|c}
\hline
\hline
\multirow{3}{*}{System} & \multirow{3}{*}{Model} & \multicolumn{2}{|c}{Fitting RMSEs} \\ \cline{3-4}
& & energy, & force, \\
& & meV/atom & meV/\AA \\ \hline
\multirow{3}{*}{$\rm CH_3COO^-$+4-methylphenol}
& MTP & 0.45 $\pm$ 0.02 & 22.0 $\pm$ 1.9 \\ \cline{2-4}
& MTP+QRd & 0.52 $\pm$ 0.16 & 23 $\pm$ 3 \\ \cline{2-4}
& MTP+EDQ & 0.13 $\pm$ 0.05 & 14.8 $\pm$ 0.8 \\ \hline
\hline
\multirow{3}{*}{$\rm CH_3COO^-$+4-methylimidazole}
& MTP & 0.731 $\pm$ 0.009 & 16.4 $\pm$ 0.8 \\ \cline{2-4}
& MTP+QRd & 0.719 $\pm$ 0.017 & 15.9 $\pm$ 0.4 \\ \cline{2-4}
& MTP+EDQ & 0.078 $\pm$ 0.005 & 10.9 $\pm$ 0.2 \\ \hline 
\hline
\end{tabular}
\end{center}
\end{table*}

We see that for both systems MTP+EDQ gives three to nine times smaller energy fitting errors than both MTP and MTP+QRd, even with such a simple charge prediction model as MTP of the 2-nd level. Furthermore, MTP+EDQ reduced force fitting errors by 30\%.

\begin{figure}[!ht]
    \centering
    \begin{subfigure}[t]{1.0\linewidth}
        \centering
        \includegraphics[width=\textwidth]{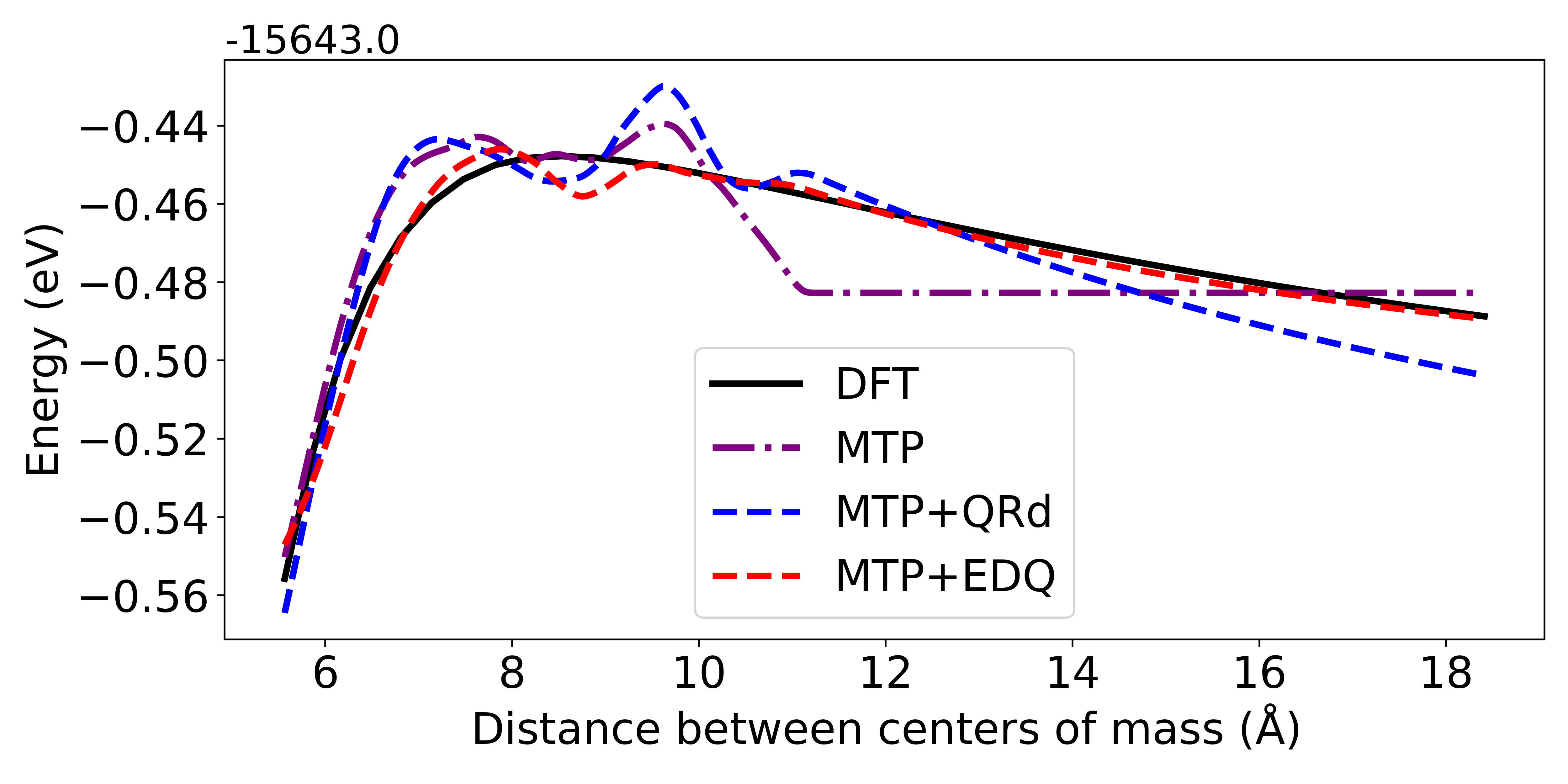}
        \caption{$\rm CH_3COO^-$+4-methylphenol.}
        \label{fig:binding_curve_phenol}
    \end{subfigure}\hfill

    \vspace{0.5cm}
    
    \begin{subfigure}[t]{1.0\linewidth}
        \centering
        \includegraphics[width=\textwidth]{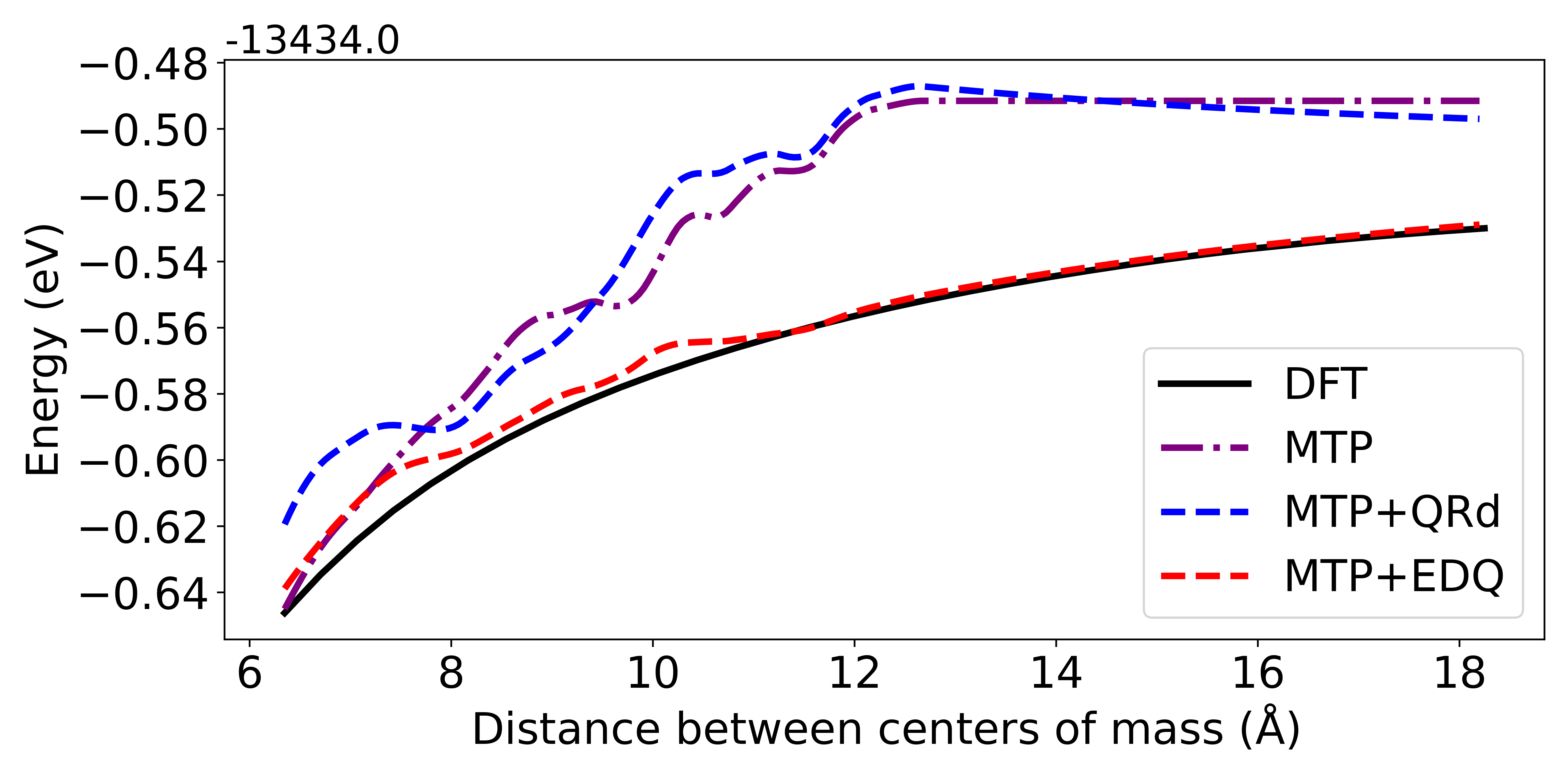}
        \caption{$\rm CH_3COO^-$+4-methylimidazole.}
        \label{fig:binding_curve_imidazole}
    \end{subfigure}
    \caption{Binding curves calculated using DFT and the ensembles of MLIPs for the $\rm CH_3COO^-$+4-methylphenol~(a) and $\rm CH_3COO^-$+4-methylimidazole~(b) systems.}
    \label{fig:binding_curves}
\end{figure}

The binding curves averaged over the ensembles of the MLIPs and obtained using DFT are given in Fig.~\ref{fig:binding_curves}.

MTP+EDQ demonstrated an ideal correspondence with the DFT binding curves obtained at large separations of molecules for both systems and a great one at small separations. Moreover, unlike MTP+QRd, the MTP+EDQ model successfully predicted the binding curve for $\rm CH_3COO^-$+4-methylimidazole qualitatively and quantitatively.

\subsection{Properties of NaCl}

In addition to non-periodic molecular systems, we verified our models on the periodic NaCl. The configurations for a training set were selected during molecular dynamics (MD) simulations using the active learning algorithm described above. MD simulations of 100 ps were conducted in the NPT-ensemble for 216 atoms at $T=300$ K and $p=1$ bar. After active learning during MD, we had 304 configurations. We utilized DFT implemented in the CP2K code~\cite{CP2K} to calculate energies, forces, and stresses of the selected configurations. We used the GTH-PBE pseudopotential with TZVP-MOLOPT-PBE-GTH basis set and a 2 $\times$ 2 $\times$ 2 k-point grid. A multi-grid contained four levels with the cutoff of the finest grid equal to 600~Ry and 50~Ry cutoff for the reference grid.

We trained ensembles of five MTPs of the 12-th and 16-th levels (MTP\textnonbreakinghyphen12 and MTP\textnonbreakinghyphen16), five MTP+EDQRd models with MTP of 12-th level and EDQRd constructed on the basis of MTP of 6-th level (MTP\textnonbreakinghyphen12+EDQRd(MTP\textnonbreakinghyphen6)), and five MTP+EDQRd models with MTP of 16-th level and EDQRd constructed on the basis of MTP of 10-th level (MTP\textnonbreakinghyphen16+EDQRd(MTP\textnonbreakinghyphen10)). All the local models, including those used for the prediction of charges, had a cutoff radius of 5~\AA \;and eight radial basis functions.

To fit the MTP+EDQRd model, we first trained MTP+EDQ by minimizing~\eqref{eq:lossfunction} with respect to the parameters of both short-range and electrostatic parts for 2000 BFGS steps, and then we fitted MTP+EDQRd by minimizing the same loss function, starting from the parameters obtained after MTP+EDQ training. We used $w_e = \frac{1.0}{216} \; \left(\text{eV}\right)^{-2}, \;w_f = 0.01 \; \left(\rm eV/\rm \AA\right)^{-2}, \;w_s = \frac{0.001}{216} \; \left(\text{eV}\right)^{-2}$ during the training of both short-range and long-range models.

The fitting errors for the ensembles of MTP\textnonbreakinghyphen12, MTP\textnonbreakinghyphen16, MTP\textnonbreakinghyphen12+EDQRd(MTP\textnonbreakinghyphen6), and MTP\textnonbreakinghyphen16+EDQRd(MTP\textnonbreakinghyphen10) are given within 1-$\sigma$ confidence interval in Table~\ref{Table:nacl_errors}.

\begin{table*}[!ht]
\caption{Fitting errors for NaCl. The results are given for the ensemble of five models with 1-$\sigma$ confidence interval.}
\label{Table:nacl_errors}
\begin{center}
\begin{tabular}{c|c|c|c}
\hline
\hline
\multirow{3}{*}{Model} & \multicolumn{3}{|c}{Fitting RMSEs} \\ \cline{2-4}
& energy, & force, & stress, \\
& $10^{-5}$ eV/atom & meV/\AA & eV \\ \hline
MTP-12 & 12.1 $\pm$ 0.5 & 12.9 $\pm$ 0.2 & 0.28 $\pm$ 0.04 \\ \hline
MTP-16 & 11.6 $\pm$ 0.4 & 12.17 $\pm$ 0.10 & 0.27 $\pm$ 0.04 \\ \hline
MTP-12+EDQRd(MTP-6) & 3.6 $\pm$ 0.5 & 2.69 $\pm$ 0.03 & 0.091 $\pm$ 0.011 \\ \hline
MTP-16+EDQRd(MTP-10) & 2.5 $\pm$ 0.5 & 2.2 $\pm$ 0.4 & 0.066 $\pm$ 0.014 \\ \hline
\hline
\end{tabular}
\end{center}
\end{table*}

From this table, we conclude that increasing the level of MTP in both short-range and electrostatic parts did not give a significant reduction of the fitting errors for both pure short-range and long-range models. Therefore, for further comparison, we chose MTP-12 and MTP-12+EDQRd(MTP-6), which we further denote by MTP and MTP+EDQRd, respectively.

\begin{figure*}[!ht]
    \centering
    \begin{subfigure}[t]{0.5\linewidth}
        \centering
        \includegraphics[width=\textwidth]{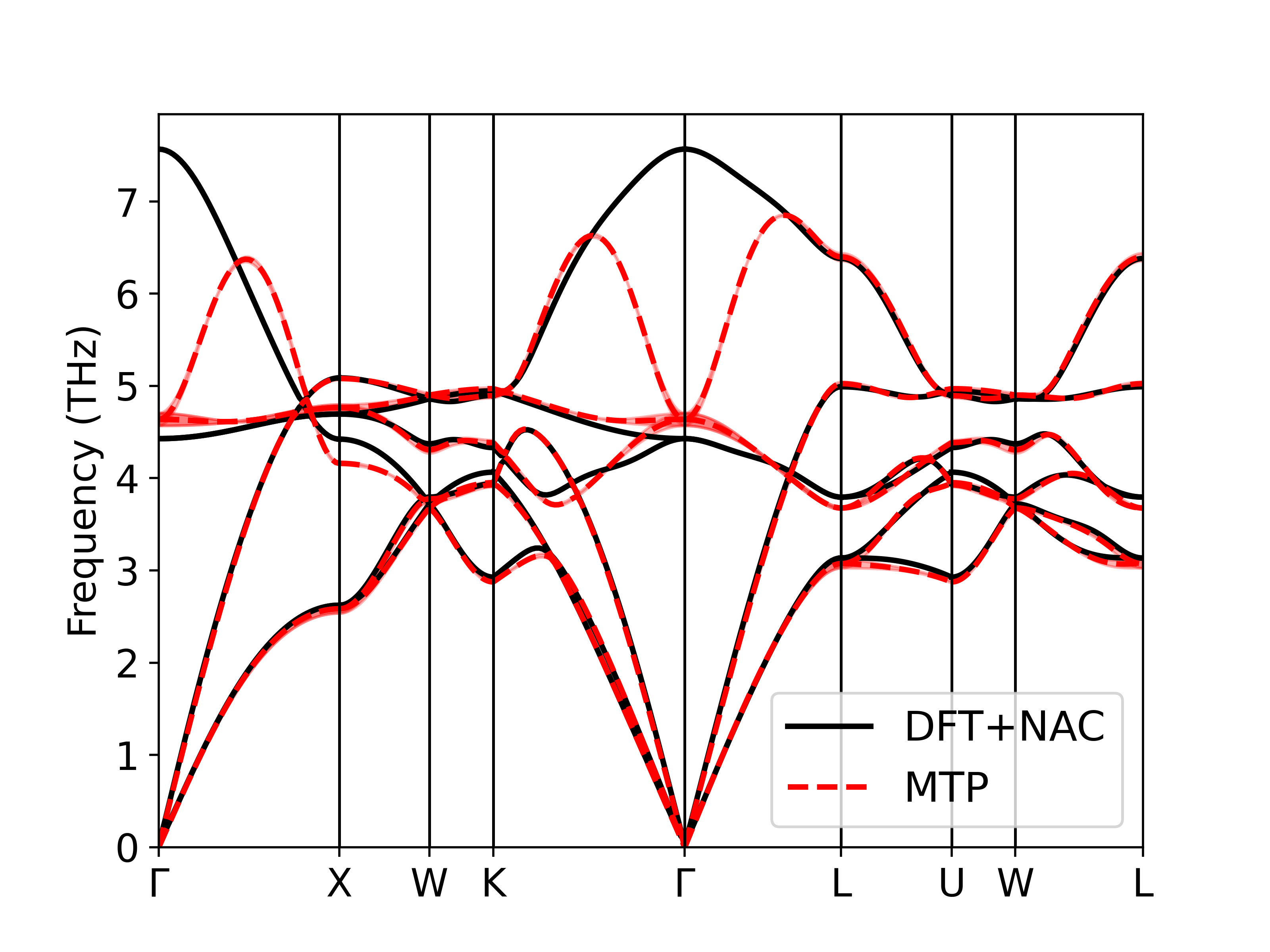}
        \caption{MTP vs DFT+NAC.}
        \label{fig:phonons_mtp_dft}
    \end{subfigure}\hfill
    \begin{subfigure}[t]{0.5\linewidth}
        \centering
        \includegraphics[width=\textwidth]{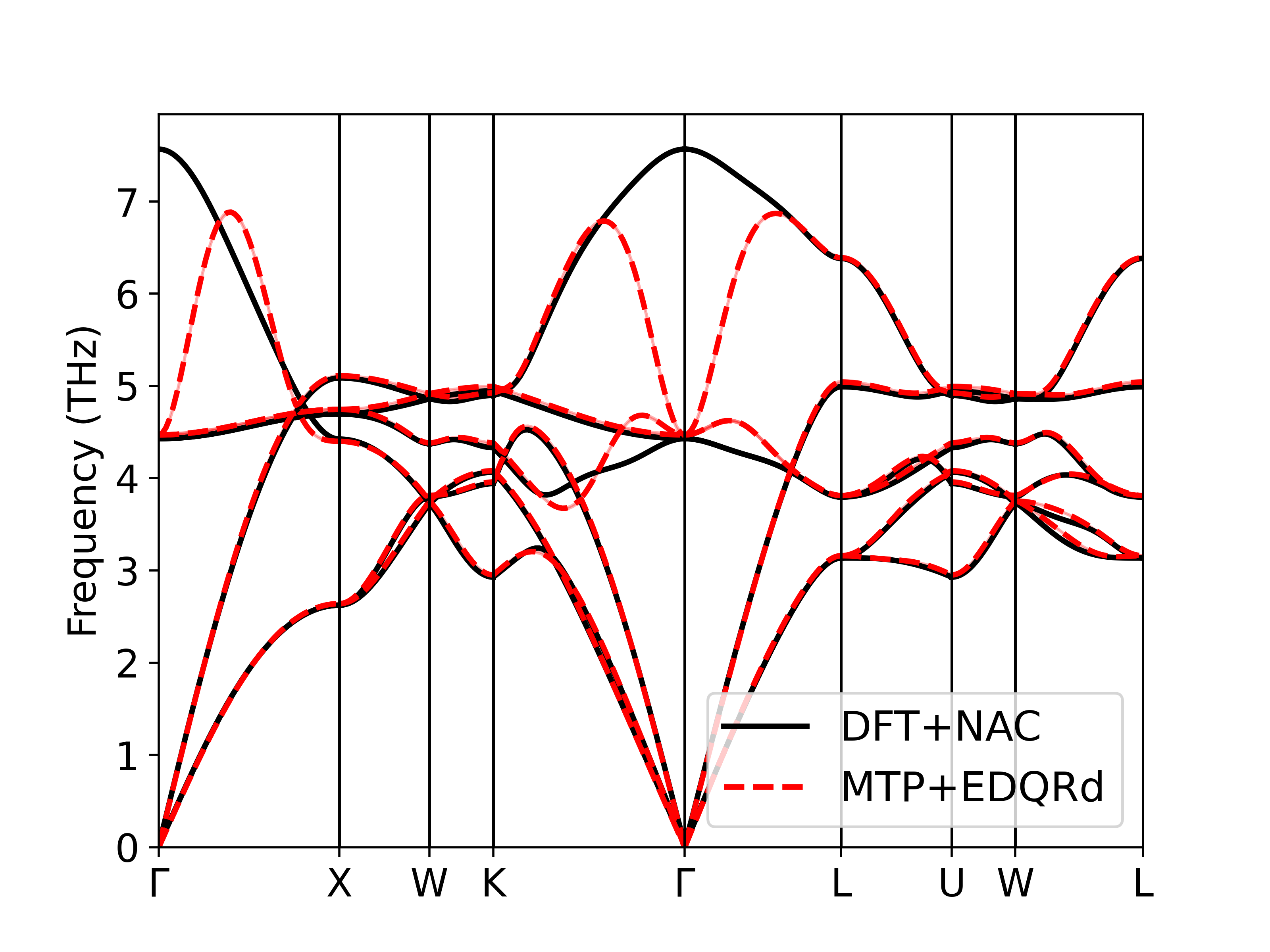}
        \caption{MTP+EDQRd vs DFT+NAC.}
        \label{fig:phonons_edqrd_dft}
    \end{subfigure}

    \vspace{0.25cm}

    \begin{subfigure}[t]{0.5\linewidth}
        \centering
        \includegraphics[width=\textwidth]{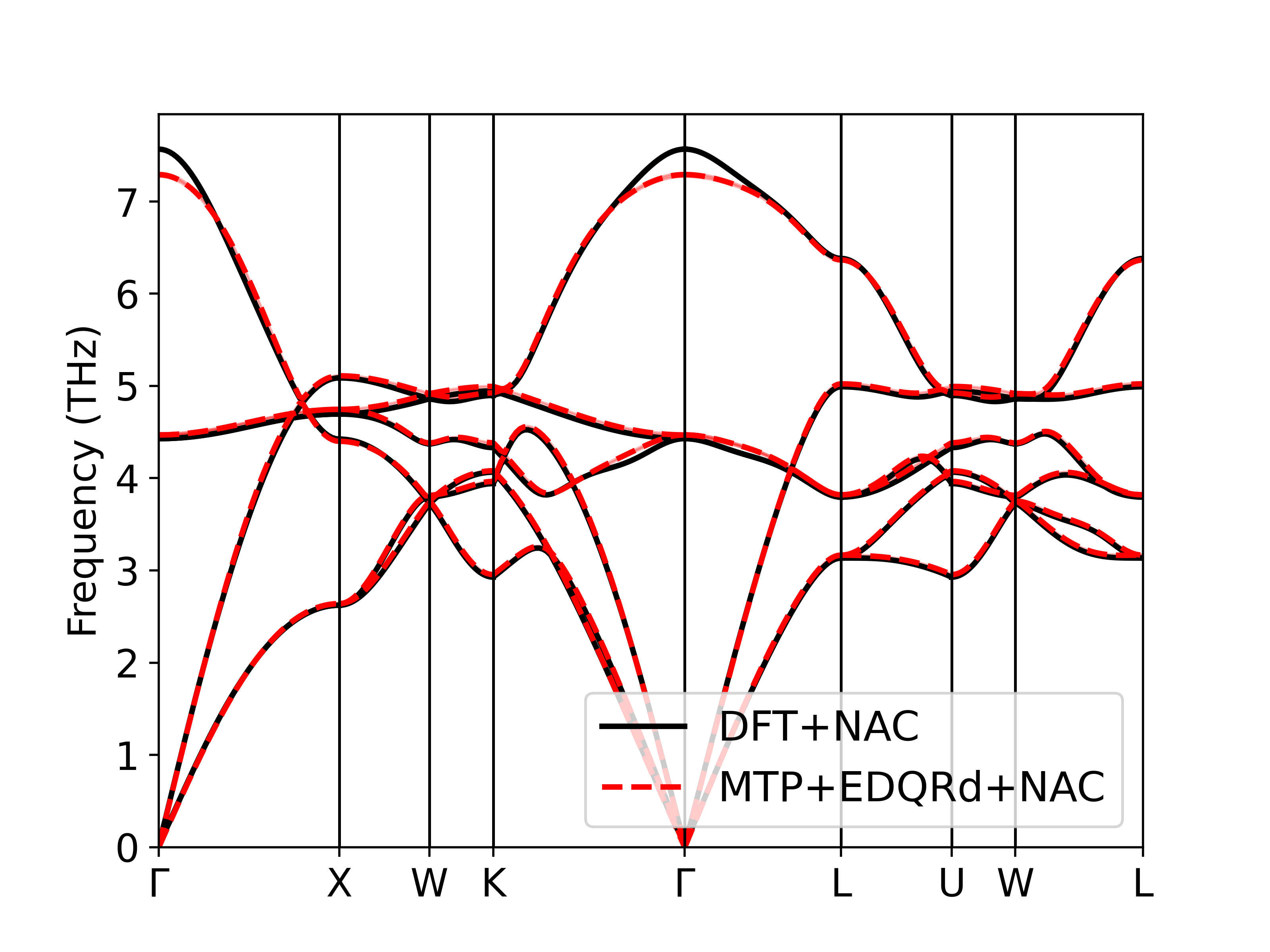}
        \caption{MTP+EDQRd+NAC vs DFT+NAC.}
        \label{fig:phonons_edqrd_nac_dft}
    \end{subfigure}\hfill
    \begin{subfigure}[t]{0.5\linewidth}
        \centering
        \includegraphics[width=\textwidth]{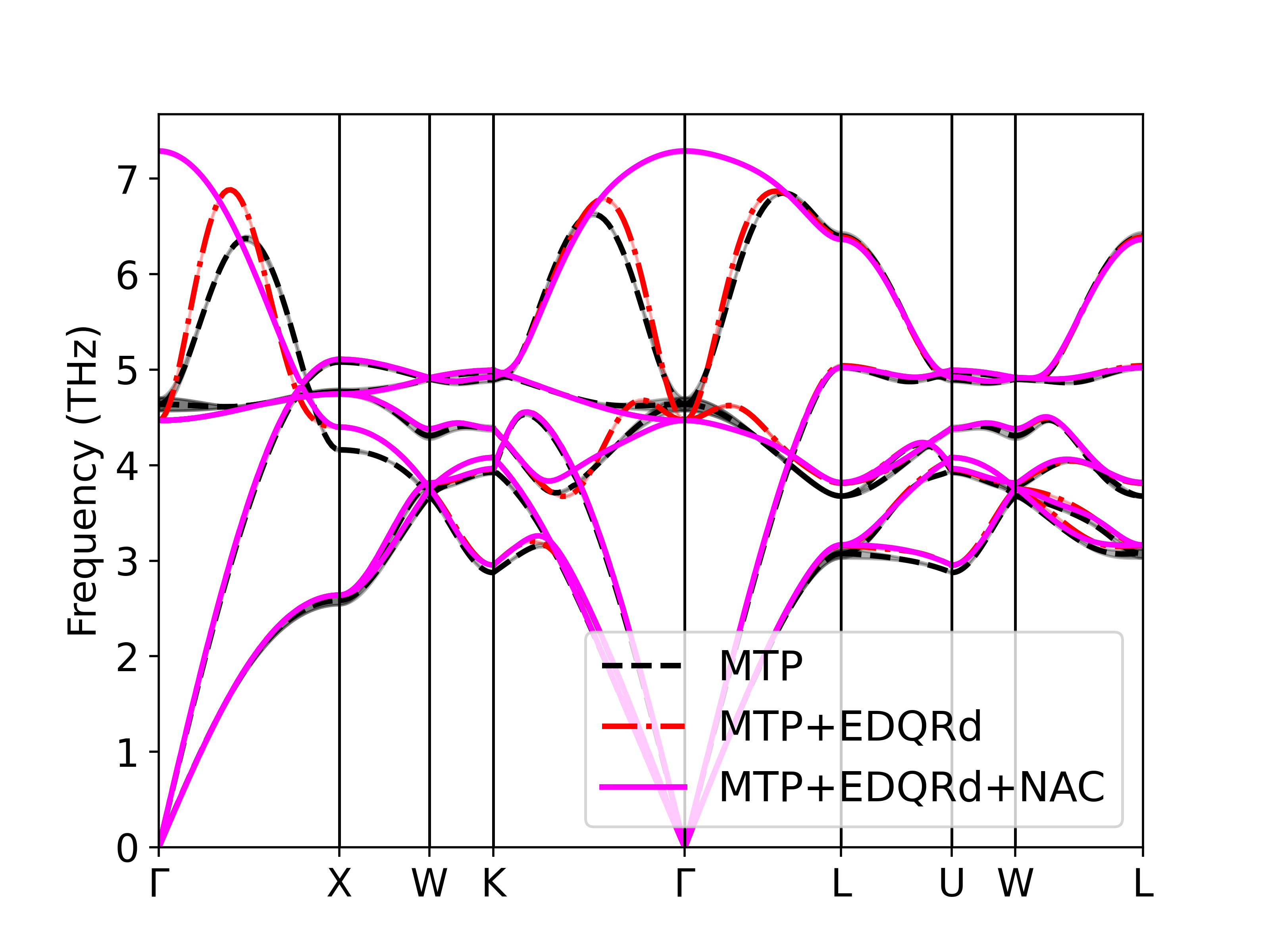}
        \caption{MTP vs MTP+EDQRd vs MTP+EDQRd+NAC.}
        \label{fig:phonons_all_mlips}
    \end{subfigure}
    
    \caption{Phonon spectra of NaCl calculated using DFT and the ensembles of MLIPs trained on energies, forces, and stresses. Shading corresponds to 1-$\sigma$ confidence interval.}
    \label{fig:phonons}
\end{figure*}

\begin{table*}[!ht]
\caption{Phonon frequencies $\Gamma_{\rm TO}$ and $\Gamma_{\rm LO}$ in the $\Gamma$-point calculated with MTP+EDQRd+NAC and DFT+NAC and scaled Born effective charges $Z^0_{\rm{Na}, \alpha \alpha}$ and $Z^0_{\rm{Cl}, \alpha \alpha}$, $\alpha=1,2,3$ obtained with MTP+EDQRd and experiment~\cite{NaCl_diel_const_ratio, NaCl_BEC}. The results for MTP+EDQRd are given for the ensemble of five models with 1-$\sigma$ confidence interval. The ensemble of MTP+EDQRd demonstrates a good agreement with DFT/experimental results.}
\label{Table:freq_and_BEC}
\begin{center}
\begin{tabular}{c|c|c|c|c}
\hline
\hline
Method & $\Gamma_{\rm TO}$, THz & $\Gamma_{\rm LO}$, THz & $Z^0_{\rm{Na}, \alpha \alpha}, e$ & $Z^0_{\rm{Cl}, \alpha \alpha}, ~e$\\ \hline
MTP+EDQRd & $4.47 \pm 0.01$ & $7.29 \pm 0.02$ & $0.695 \pm 0.004$ & $-0.682 \pm 0.003$ \\ \hline
DFT/experiment & $4.43$ & $7.57$ & $0.734$ & $-0.734$ \\ \hline
\hline
\end{tabular}
\end{center}
\end{table*}

The MTP+EDQRd model yielded three times smaller energy and stress fitting errors and five times smaller force fitting error than MTP. Since all potentials in one ensemble produced approximately the same errors, we randomly chose one representative of the ensemble and computed the lattice constant at zero temperature and density at a temperature of 300~K and a pressure of 1~bar only for those representatives. All of our calculations (including DFT) yielded the same lattice constants of 5.66~\AA \;, which indicates that all the models learned to predict stresses correctly. To obtain the density, we conducted 300~ps of MD simulations of a supercell with 1728 atoms in the NpT-ensemble using LAMMPS~\cite{LAMMPS} and averaged the density of the system over the last 200~ps. For both MTP and MTP+EDQRd, we obtained the density value of $2.05 \; \rm g/cm^3$, which differs by 5\% from the experimental value of $2.156 \pm 0.001 \; \rm g/cm^3$~\cite{NaCl_exp_density}.

Then, we obtained phonon spectra using four different methods. Firstly, we used DFT with the non-analytical correction~\eqref{eq:nac} calculated using high-frequency dielectric constant $\varepsilon_{\infty}=2.33$~\cite{NaCl_diel_const_ratio} and BECs $Z^*_{{\rm Na},\alpha \alpha}=1.12 ~e$ and $Z^*_{{\rm Cl},\alpha \alpha}=-1.12 ~e$, $\alpha=1,2,3$~\cite{NaCl_BEC} taken from experiments (these results are denoted as DFT+NAC). Then, we utilized MTP and MTP+EDQRd without any non-analytical correction. Finally, we adopted MTP+EDQRd with the non-analytical correction~\eqref{eq:nacfinal} calculated using scaled BECs predicted by EDQRd (denoted as MTP+EDQRd+NAC). These calculations were performed using Phonopy~\cite{phonopy-phono3py-JPCM}, and to avoid finite-size effects, the scaled BECs needed to obtain NAC for MTP+EDQRd were calculated in a $10\times10\times10$ supercell (8000 atoms), while the calculations of the force constants were done in a $3\times3\times3$ supercell (216 atoms). Phonon spectra averaged over ensembles of five MLIPs and the one obtained using DFT are shown in Fig.~\ref{fig:phonons}.

All the models yielded the results corresponding to DFT+NAC except for the vicinity of the $\Gamma$-point. There are several notable differences in its vicinity.
\begin{itemize}
    \item MTP merged some branches on the U-W-K path and overestimated the frequency of the transverse optical~(TO) modes at the $\Gamma$-point.
    \item MTP+EDQRd also lacked the LO-TO splitting, but correctly predicted the frequency of the TO modes at the $\Gamma$-point and improved the prediction at X and L. However, it could not give the correct frequencies of TO in the vicinity of the $\Gamma$-point (not in the point itself).
    \item MTP+EDQRd+NAC yielded a spectrum that was in excellent agreement with the DFT+NAC one along the k-path and slightly underestimated the value of the LO-TO splitting, which illustrates that the proposed method of the phonon spectrum calculation for isotropic materials is valid.
\end{itemize}

The slight discrepancy between the predictions of MTP+EDQRd+NAC and DFT+NAC can, possibly, be attributed to the fact that we used experimental values of BECs and high-frequency dielectric constant to compute NAC for DFT.

In addition, we provide and compare phonon frequencies at the $\Gamma$-point calculated with MTP+EDQRd+NAC and DFT+NAC as well as scaled BECs obtained with MTP+EDQRd and experiment~\cite{NaCl_diel_const_ratio, NaCl_BEC} in Table~\ref{Table:freq_and_BEC}. We see that all the physical quantities calculated with MTP+EDQRd are close either to DFT (phonon frequencies) or to experiment (scaled BECs), which demonstrates a good predictive power of the developed MTP+EDQRd model.

\begin{figure}[!ht]
    \centering
    \includegraphics[width=\linewidth]{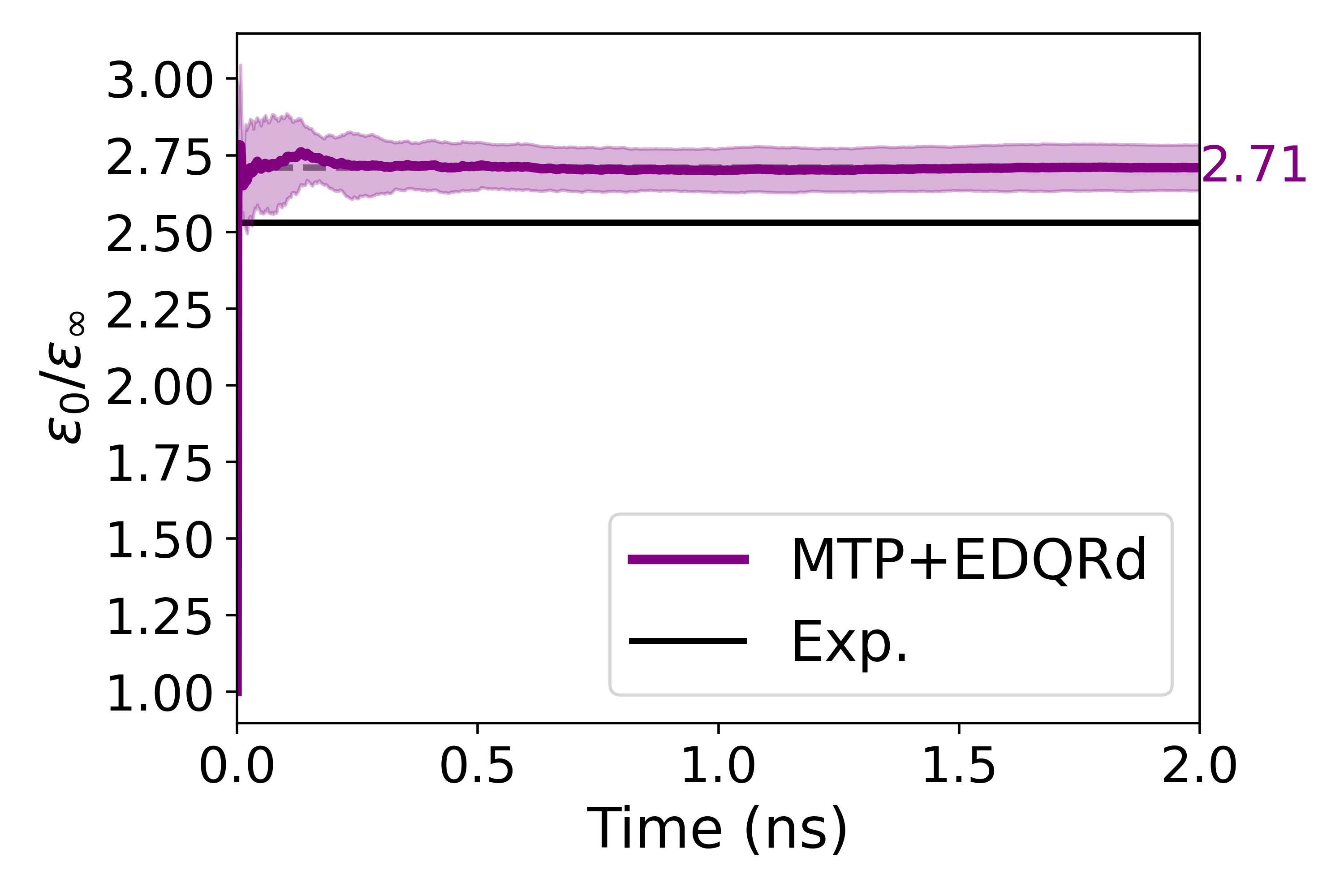}
    \caption{The ratio of the static and high-frequency dielectric constants in NaCl calculated using MTP+EDQRd. Shading corresponds to the 1-$\sigma$ confidence interval, experimental value were taken from~\cite{NaCl_diel_const_ratio}.}
    \label{fig:diel_consts}
\end{figure}

Finally, we calculated the ratio of the static and high-frequency dielectric constants for NaCl. To obtain the averaged value of dipole moment required for this calculation, we ran the MD simulation in LAMMPS~\cite{LAMMPS}. We took a non-equilibrium cell with 5832~atoms and equilibrated it firstly in the NpT ensemble at 290~K and 1~bar for 50~ps, and then in the NVT ensemble at 290~K for 50~ps. Then, we ran a 2~ns simulation in the NVT ensemble using Bussi-Donadio-Parrinello thermostat~\cite{BDPThermostat} at 290~K, during each time step of which we calculated the dipole moment and used it to obtain the ratio of dielectric constants. The calculation was done for all the five MTP+EDQRd models in the ensemble and the results were averaged. The ratio of the dielectric constants as a function of the simulation time together within 1-$\sigma$ confidence interval is shown in Fig.~\ref{fig:diel_consts}.

The average and the standard deviation of the ratio of interest converged after 0.5~ns, and the standard deviation does not change during the simulation time after 0.5~ns indicating that our models yielded slightly different predictions. We obtained the value of $\varepsilon_0/\varepsilon_\infty$ of $2.71 \pm 0.07$, which is in an excellent agreement with the experiment: it is only 7\% higher than the experimental value of 2.53~\cite{NaCl_diel_const_ratio}.

\subsection{Phonon spectrum of PbTiO$_3$}

Although \eqref{eq:nacfinal} is strictly valid only for isotropic materials, here we investigate the applicability of~\eqref{eq:nacfinal} to uniaxial materials and use it for tetragonal PbTiO$_3$, in which ordinary and extraordinary high-frequency dielectric constants differ by 10\%~\cite{PbTiO3_diel_const}.

We trained an MTP+EDQRd with MTP of the 20-th level and EDQRd constructed on the basis of MTP of the 16-th level. Both local models had a cutoff radius of 5~\AA \;and eight radial basis functions. We used the training and validation sets utilized in~\cite{LES_BEC} and constructed in~\cite{PbTiO3_training_set}. These sets contain configurations with energies and forces of PbTiO$_3$ in both cubic and tetragonal phases, which were calculated using the SCAN density functional~\cite{SCAN}. The training and validation sets contained 3545 and 600 configurations with 135 atoms each, respectively. To train the model, we utilized the same procedure as for NaCl with the training weights $w_e = \frac{1.0}{135} \; \left(\text{eV}\right)^{-2}, \;w_f = 0.01 \; \left(\rm eV/\rm \AA\right)^{-2}, \;w_s = 0 \; \left(\text{eV}\right)^{-2}$. MTP+EDQRd yielded the validation errors of 0.6~meV/atom in energy and 91.8 meV/\AA \;in forces, which are close to the 0.4~meV/atom and 79.8 meV/\AA \;reported for the CACE+LES model in~\cite{LES_BEC}.

\begin{figure}[!ht]
    \centering
    \begin{subfigure}[t]{1.0\linewidth}
        \centering
        \includegraphics[width=\textwidth]{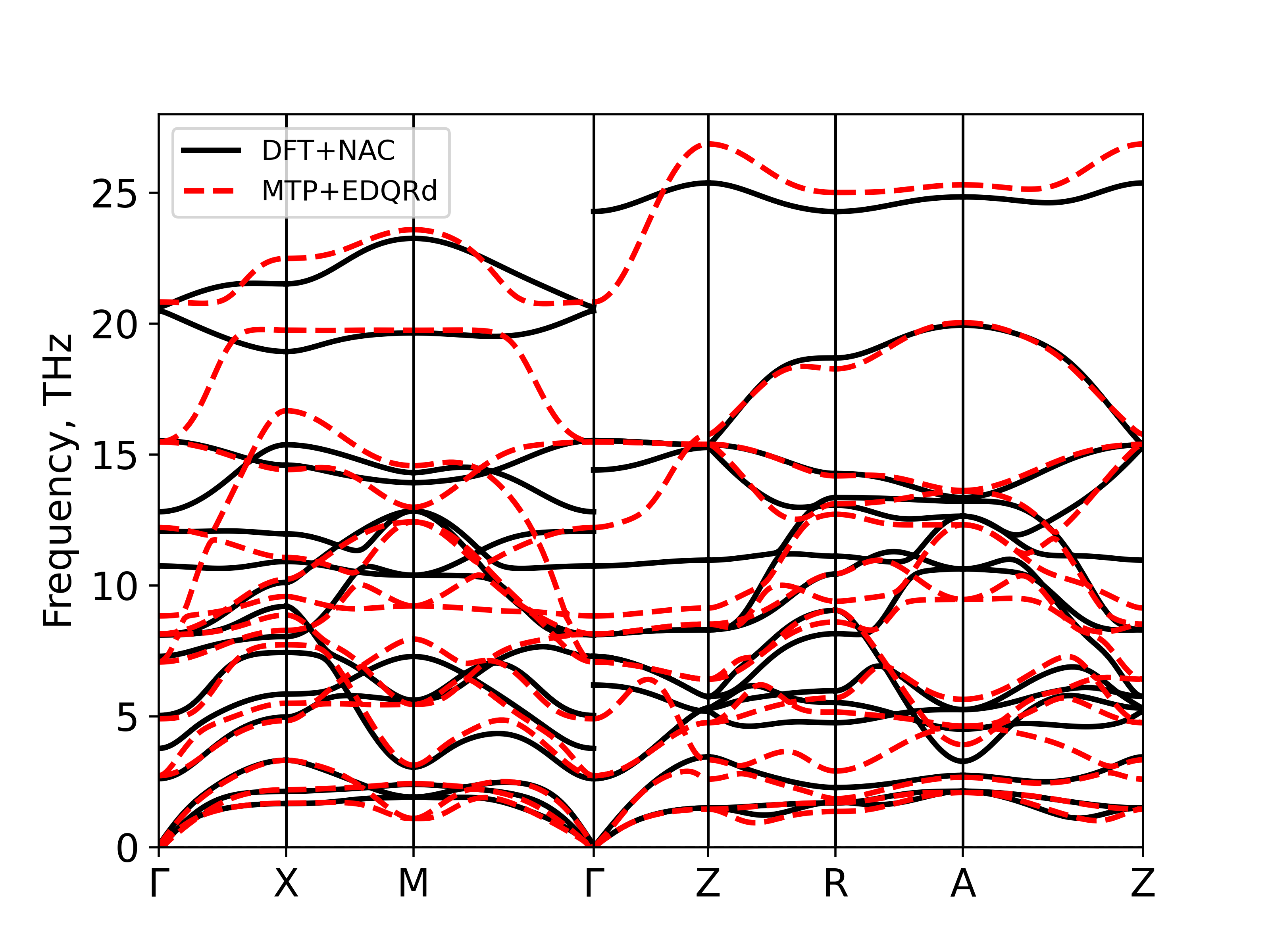}
        \caption{MTP+EDQRd vs DFT+NAC.}
        \label{fig:pbtio3_phonons_edqrd}
    \end{subfigure}\hfill

    \vspace{0.5cm}
    
    \begin{subfigure}[t]{1.0\linewidth}
        \centering
        \includegraphics[width=\textwidth]{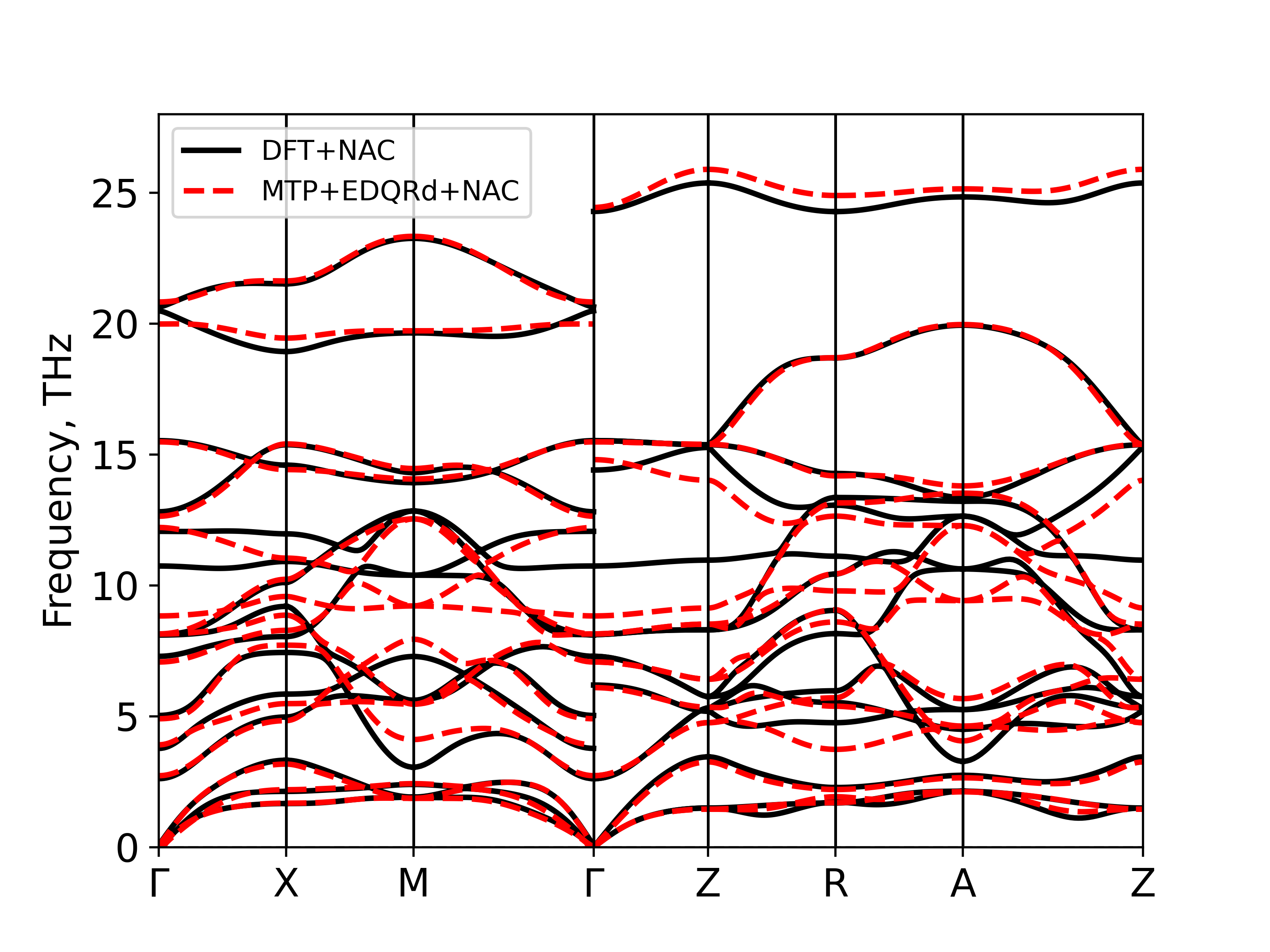}
        \caption{MTP+EDQRd+NAC vs DFT+NAC.}
        \label{fig:pbtio3_phonons_edqrd_nac}
    \end{subfigure}
    \caption{Phonon spectrum of PbTiO$_3$ calculated using DFT+NAC and MTP+EDQRd without and with NAC trained on energies and forces. MTP+EDQRd+NAC phonon spectrum is overall in excellent agreement with the DFT+NAC.}
    \label{fig:pbtio3_phonons}
\end{figure}

We further used the trained MTP+EDQRd model to calculate the phonon spectrum of the tetragonal PbTiO$_3$ both without and with NAC. To avoid finite-size effects, the BECs were calculated in $12\times12\times12$ supercell (8640 atoms), while the calculation of the force constants was carried out in $3\times3\times3$ supercell (135 atoms). For comparison with MTP+EDQRd, we also calculated the PbTiO$_3$ phonon spectrum with DFT+NAC using the same supercell of 135 atoms as for MTP+EDQRd and the same parameters that were utilized for the training set calculation: the energy cutoff of 150 Ry and the $1 \times 1 \times 1$ k-mesh~\cite{PbTiO3_training_set}. The tensors of high-frequency dielectric constants and BECs were taken from \cite{PbTiO3_dft_phonons}. The comparison of the spectra is given in Fig.~\ref{fig:pbtio3_phonons}.

MTP+EDQRd yielded spectra that is in a qualitative agreement with the DFT+NAC one far from the $\Gamma$-point, while MTP+EDQRd+NAC yielded an excellent quantitative agreement for almost all modes. Some details are given below.
\begin{itemize} 
    \item Acoustic modes and high-frequency part of the spectrum $(\geq 15 \; \rm THz)$ predicted by MTP+EDQRd+NAC are in an excellent agreement with the DFT+NAC one, while MTP+EDQRd significantly underestimates frequencies of two acoustic modes in the M-point and gives good quantitative predictions of the high-frequency modes only far from the $\Gamma$-point.
    \item MTP+EDQRd and MTP+EDQRd+NAC produced close optical modes with frequencies smaller than 15~THz far from the $\Gamma$-point, with NAC mostly flatting out the curves. Notably, our models give a 2~THz error in an almost horizontal line at 12~THz.
    \item MTP+EDQRd+NAC yielded spectrum that is in an excellent agreement with the DFT+NAC one in the vicinity of the $\Gamma$-point, except for the aforementioned error of 2~THz for one mode.
\end{itemize}

To analyze the performance of our NAC calculated using~\eqref{eq:nacfinal}, we compared the frequency shifts in the small vicinity of the $\Gamma$-point introduced by the addition of NAC given by~\eqref{eq:nac} to DFT and frequency shifts given by our NAC. Obtained results are shown in Table~\ref{Table:freq_shifts}.

\begin{table}[!ht]
\caption{Frequency shifts introduced by NAC in the small vicinity of the $\Gamma$-point. They were calculated along three directions: $\Gamma\rightarrow\rm X$, $\rm M\rightarrow \Gamma$, and $\Gamma\rightarrow\rm Z$. The shifts predicted with MTP+EDRQd are mostly in excellent agreement with DFT.}
\label{Table:freq_shifts}
\begin{center}
\begin{tabular}{c|c|c|c|c||c|c|c}
\hline
\hline
\multirow{2}{*}{Model} & \multicolumn{7}{c}{Frequency shifts, THz} \\ \cline{2-8}
& \multicolumn{4}{c||}{$\Gamma\rightarrow\rm X$ and $\rm M\rightarrow \Gamma$} & \multicolumn{3}{c}{$\Gamma\rightarrow\rm Z$} \\ \hline
DFT & 1.16 & 5.53 & -0.018 & 5.11 & 1.16 & 2.34 & 3.80 \\ \hline
MTP+EDQRd & 1.18 & 5.57 & -0.004 & 4.50 & 1.19 & 2.59 & 3.61 \\ \hline
\hline
\end{tabular}
\end{center}
\end{table}

We see that the calculated frequency shifts are in an excellent agreement, with the biggest discrepancy of 0.61 THz or 12\% (excluding one point with extremely small shift). Overall, we conclude that even though~\eqref{eq:nacfinal} is not strictly valid for anisotropic materials, it may remain useful beyond the isotropic case and can significantly improve the spectrum obtained without NAC.

\section{Conclusion}

In this study, we proposed two machine-learning interatomic potentials (MLIPs) with explicit long-range electrostatics in the form of the Coulomb interaction of point charges and introduced a method for calculating phonon spectra of isotropic materials. The new Environment-Dependent Charge Redistribution (EDQRd) model and its simplified version without charge redistribution (Environment-Dependent Charges, EDQ) were combined with one of the short-range MLIPs, namely, Moment Tensor Potential (MTP), which was used in the prediction of both short-range part of the energy and of the atomic charges depending on their environments. 

First, we tested the long-range MTP+EDQ model on organic dimers in vacuum (non-periodic systems) and compared the results with short-range MTP and with the long-range MTP+QRd model including fixed charges, which was recently proposed in~\cite{korogod2025_mtp_coulomb_fixed}. MTP+EDQ yielded an energy root-mean square error (RMSE) that was at least three times smaller than that given by both MTP and MTP+QRd. Furthermore, MTP+EDQ was able to quantitatively predict the binding curve in the $\rm CH_3COO^-$+4-methylimidazole system, while MTP+QRd could not even give a qualitatively-correct prediction. 

Next, we tested our potentials on the periodic NaCl crystal. To create the training set, we used active learning of short-range MTP during molecular dynamics. We further fitted MTP and MTP+EDQRd on this training set and showed that MTP+EDQRd yielded two times smaller energy and stress fitting RMSEs and five times smaller force RMSE than MTP. In spite of a discrepancy in fitting errors, MTP and MTP+EDQRd yielded identical values of density, which were close to the experimental one, and the same lattice constants corresponding to the DFT one. An important result of this work is a proposed method for calculating phonon spectra of isotropic materials. To capture LO-TO splitting in the $\Gamma$-point, we added a non-analytical correction (NAC) term to the dynamical matrix. We derived that it is not necessary to know neither high-frequency dielectric constant nor Born Effective Charges (BECs) for the NAC calculation in an isotropic material and only a specific combination of them that can be predicted via long-range MLIPs is required. This result opens up the possibility of calculating the NAC to the dynamical matrix using only long-range MLIPs fitted to energies, forces, and, possibly, stresses. We applied this method to the NaCl crystal, and obtained the phonon spectrum that was in good agreement with the DFT one with a slightly underestimated value of the LO-TO splitting. Then, we calculated the ratio of the static and high-frequency dielectric constants and got the value of $2.71 \pm 0.07$, which is only 7\% higher than 2.53 obtained in experiment~\cite{NaCl_diel_const_ratio}.

Finally, we verified our method for NAC calculation that is strictly correct only for isotropic materials on the uniaxial material --- tetragonal PbTiO$_3$. Our MTP+EDQRd model reached the accuracy of the other recently developed long-range CACE+LES model~\cite{LES_BEC}. Moreover, the addition of NAC calculated with this MTP+EDQRd potential and the proposed method drastically improved the prediction of the phonon spectrum even despite the anisotropy of PbTiO$_3$.

In our future work, we plan to develop an active learning algorithm for automated fitting of the proposed long-range models to avoid creating a training set on the basis of the short-range MTP. Another direction of our studies will be the calculation of dielectric tensors of anisotropic materials.

\section*{Data availability}

The data that support the findings of this work are publicly available~\cite{datasets_and_trained_pots}.

\section*{Acknowledgments}

The work was supported by the grant for research centers in the field of AI provided by the Ministry of Economic Development of the Russian Federation in accordance with the agreement 000000C313925P4E0002 and the agreement with HSE University No 139-15-2025-009. This research was supported in part by computational resources of HPC facilities at the HSE University~\cite{kostenetskiy2021hpc}.

\bibliography{refs}

\end{document}